\documentstyle[11pt,moriond]{article}

\def\be{\begin{equation}}
\def\ee{\end{equation}}
\def\bea{\begin{eqnarray}}
\def\eea{\end{eqnarray}}

\def\ltsima{$\; \buildrel < \over \sim \;$}
\def\lsim{\lower.5ex\hbox{\ltsima}}
\def\gtsima{$\; \buildrel > \over \sim \;$}
\def\gsim{\lower.5ex\hbox{\gtsima}}

\begin{document}
\vspace*{4cm}
\title{NUMERICAL INVESTIGATIONS OF WEAK LENSING BY LARGE-SCALE STRUCTURE}

\author{Bhuvnesh Jain $^1$, Uro\v s Seljak $^2$, Simon White $^3$}

\address{$^{1}$
Dept. of Physics, Johns Hopkins University, Baltimore, MD 21218, USA\\
$^{2}$
Harvard Smithsonian Center For Astrophysics, Cambridge, MA 02138, USA\\
$^{3}$
Max-Planck-Institut f\"ur Astrophysik, Garching 85740, GERMANY}

\maketitle\abstracts{
We use numerical
simulations of ray tracing through N-body simulations to investigate
weak lensing by large-scale structure.
These are needed for testing the analytic predictions of two-point
correlators, to set error estimates on them and to investigate 
nonlinear gravitational effects in the weak lensing maps.
On scales larger than 1 degree gaussian statistics suffice and 
can be used to estimate the sampling, noise and aliasing errors on 
the measured power spectrum. For this case we describe a minimum 
variance inversion procedure from the 2-d to 3-d power spectrum and discuss  
a sparse sampling strategy which optimizes the signal to noise on 
the power spectrum. On degree scales and smaller the shear and convergence
statistics lie in the nonlinear regime and have a non-gaussian
distribution. For this regime ray tracing simulations are useful to 
provide reliable error estimates and calibration of the measurements. 
We show how the skewness and kurtosis can in principle be used to probe 
the mean density in the universe, but are sensitive to sampling 
errors and require large observed areas. The probability distribution 
function is likely to be more useful as a tool to investigate 
nonlinear effects. In particular, it shows striking differences 
between models with different values of the mean density $\Omega_m$. 
}

\section{Introduction}

Weak lensing by large-scale structure (LSS) shears the images of distant
galaxies. 
The first calculations of weak lensing by LSS (Blandford et al. 1991; 
Miralda-Escude 1991; Kaiser 1992), based on the pioneering work of
Gunn (1967), showed that lensing would induce coherent ellipticities
of order 1$\%$ over regions of order one degree on the sky. Recently several
authors have extended this work to probe semi-analytically the 
possibility of measuring  the mass power spectrum and cosmological parameters
from the second and third moments of the induced ellipticity or 
convergence (e.g. Bernardeau et al. 1997; Jain and Seljak 1997; Schneider 
et al. 1997).

The analytical work cited above suggested that nonlinear evolution of
the density perturbations that provide the lensing effect can significantly
alter the predicted signal. It enhances the power spectrum on scales
below one degree and makes the probability distribution function (pdf) of
the ellipticity and convergence non-Gaussian. We have carried out
numerical simulations of ray tracing through N-body simulation data to
compute the fully nonlinear moments and pdf. Details of the method
and results are presented in a forthcoming paper; here we summarize
the method and present some highlights of the results in Figures 1-5. 
We also discuss reconstruction of the dark matter power spectrum 
and error estimation using the gaussian approximation.

\section{Ray Tracing Method}

The dark matter distribution obtained from N-body simulations of
different models of structure formation is projected on to 2-dimensional
planes lying between the observer and source galaxies. Typically we 
use galaxies at $z\sim 1$ with
$\sim 20-30$ planes. We propagate $\sim 10^6$ light rays through these planes
by computing the deflections due to the matter at every plane. 
Fast Fourier Transforms are used to compute gradients of the potential
that provide the shear tensor at each plane. The outcome
of the simulation is a map of the shear and convergence on square
patches of side length $1-5^{\circ}$. Several realizations for each model
are needed to compute reliable statistics on scales ranging from $1'$
to 1$^{\circ}$. Figure 1 shows the shear map from two cold dark matter models
with shape paratemeter $\Gamma=0.2$, one with $\Omega_m=1$ (upper panel)
and the other with $\Omega_m=0.3$ (lower panel). 

Numerical resolution on small scales is limited primarily by (i) the 
force softening used in the N-body simulations, and, (ii) by the finite
size of the grid used to compute FFT's and to propagate the light
rays. We have chosen grid spacings in order to avail of the full
resolution provided by the N-body simulations. The simulations used,
adaptive $P^3M$ simulations with $256^3$ particles, were carried out
using codes kindly made available by the Virgo consortium (e.g. 
Jenkins et al. 1997) and have previously been analyzed in the strong
lensing regime by Bartelmann et al. (1998). These, 
coupled with ray tracing on a $2048^2$ grid, provide us with small scale 
resolution down to $\lsim 0.5'$, well into the nonlinear regime for weak
lensing. On large scales the finite
number of transverse modes of the density field at any redshift
leads to fluctuations that must be averaged using several realizations
of the ray tracing as well as the N-body simulation for a given model. 
We have used an ensemble of $PM$ simulations to get reliable 
statistics on large scales. 

In the weak lensing regime, the magnification and induced ellipticity
are given by linear combinations of the Jacobian matrix of the mapping
from the source to the image plane. The Jacobian matrix is defined by
\be
\Phi_{ij} \equiv  {\partial \delta \theta_i \over \partial \theta_j}
\ee
where $\delta \theta_i$ is the $i-$th component of the perturbation due to
lensing of the angular position on the 
source plane, and $\theta_j$ is the $j-$th component of the position on 
the image plane. The convergence is defined as 
$\kappa=-(\Phi_{11}+\Phi_{22})/2$, while
the two components of the shear are $\gamma_1=-(\Phi_{11}-\Phi_{22})/2$ 
and $\gamma_2=-\Phi_{12}$. The convergence $\kappa$ can be reconstructed
from the measured shear $\gamma_1$, $\gamma_2$, up to a constant which 
depends on the mean density in the survey area. If the survey is 
sufficiently large and there is little power on scales larger than the
survey, this error can be neglected.

\begin{figure}[p]
\vspace*{17.5cm}
\caption{The shear pattern on a 1$^{\circ}$ field for an
Einstein-de Sitter model (upper panel) and an open model (lower panel)
with $\Omega_m=0.3$. 
The rms shear in these fields is about $2\%$. The open model 
produces a shear pattern dominated mostly by halos of clusters or groups.
The Einstein-de Sitter model has a lower normalization ($\sigma_8=0.6$)
compared to the open model ($\sigma_8=0.85$) and evolves more rapidly 
with redshift -- this leads to a smaller fraction of the mass being
part of massive collapsed halos. The qualitative differences between
the models are reflected in the statistical measures shown in the following
figures. 
}
\includegraphics{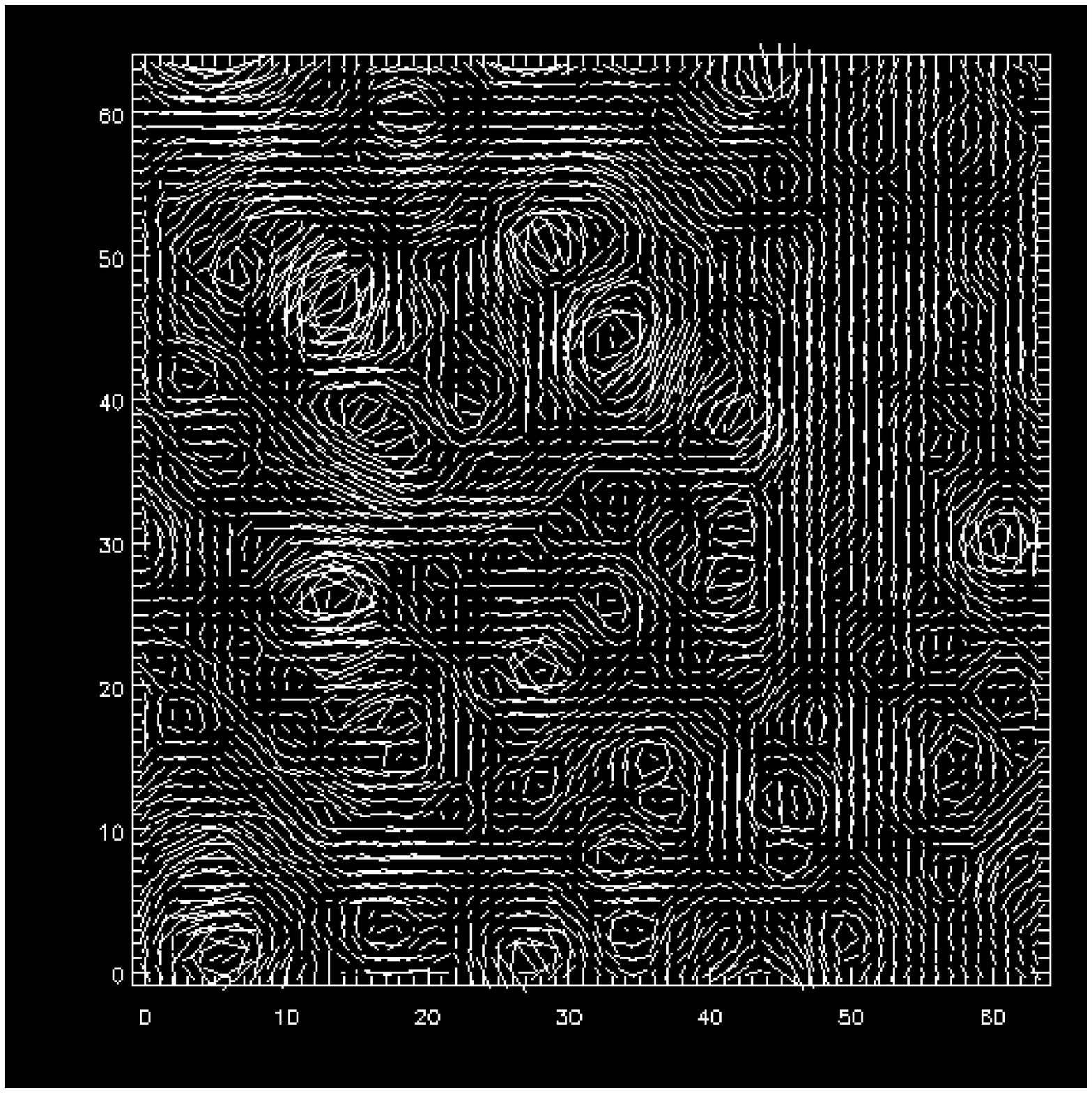}
\includegraphics{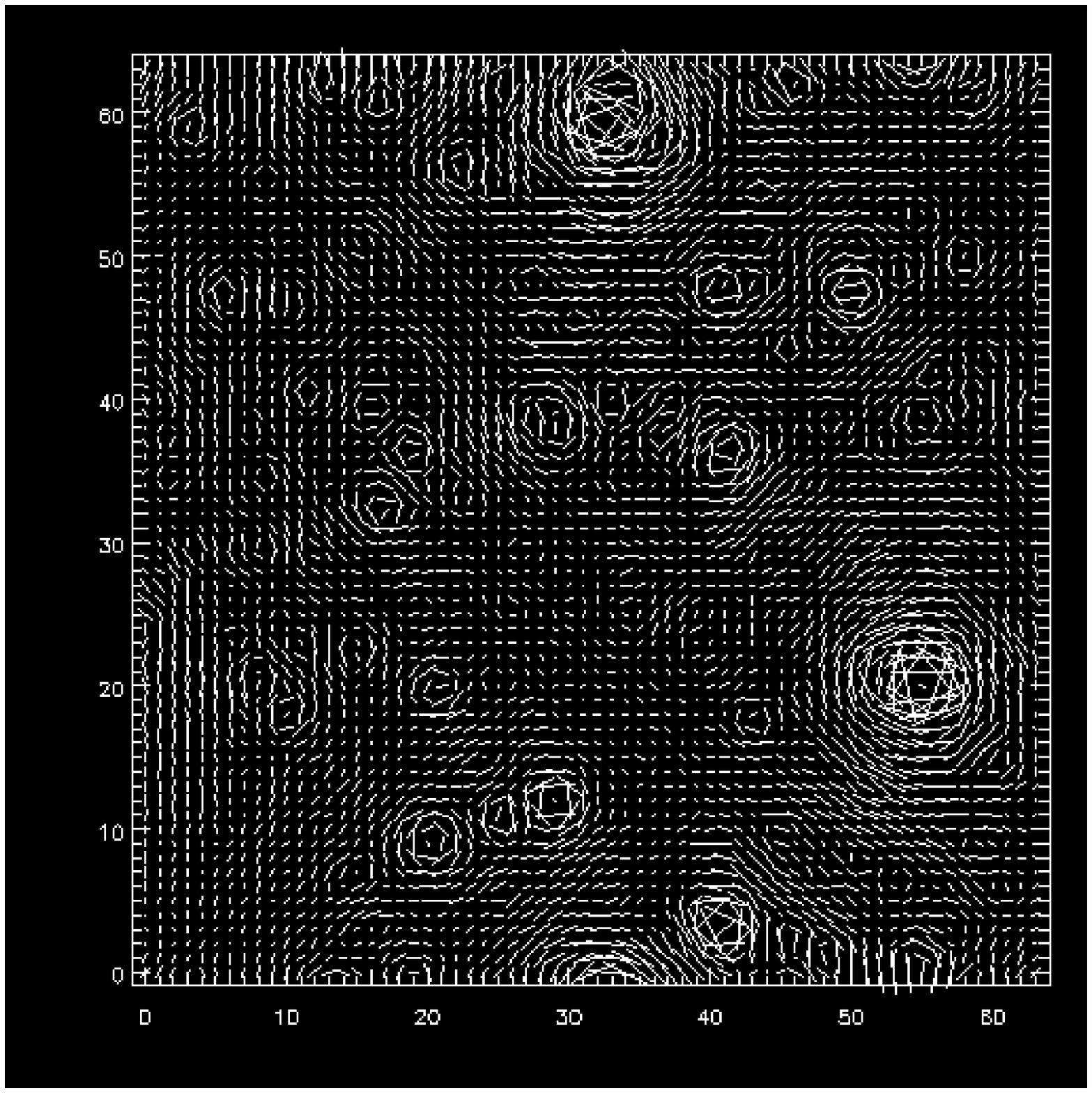}
\label{figshear}
\end{figure}

\section{Results of Simulations}

\begin{figure}[t]
\vspace*{8cm}
\caption{The dimensionless power spectrum of $\kappa$ and
$\gamma$. For the cosmological model indicated by $\Omega_m$ and
$\Gamma$ in the panel, the power spectrum from ray tracing is
compared with the linear (long-dashed) and nonlinear analytical 
(short-dashed) predictions. The dotted curves in two of the panels 
show measurements from individual realizations while the solid curves 
are the mean spectra. 
The angular wavenumber $l$ is given in inverse radians -- the smallest
$l$ plotted corresponds to modes with wavelength of order 
$L\simeq 2.5^{\circ}$, where $L$ is the side-length of the field. 
Note that the decline in the N-body power spectrum at $l>10^4$ is
because of the limited resolution.
}
\includegraphics{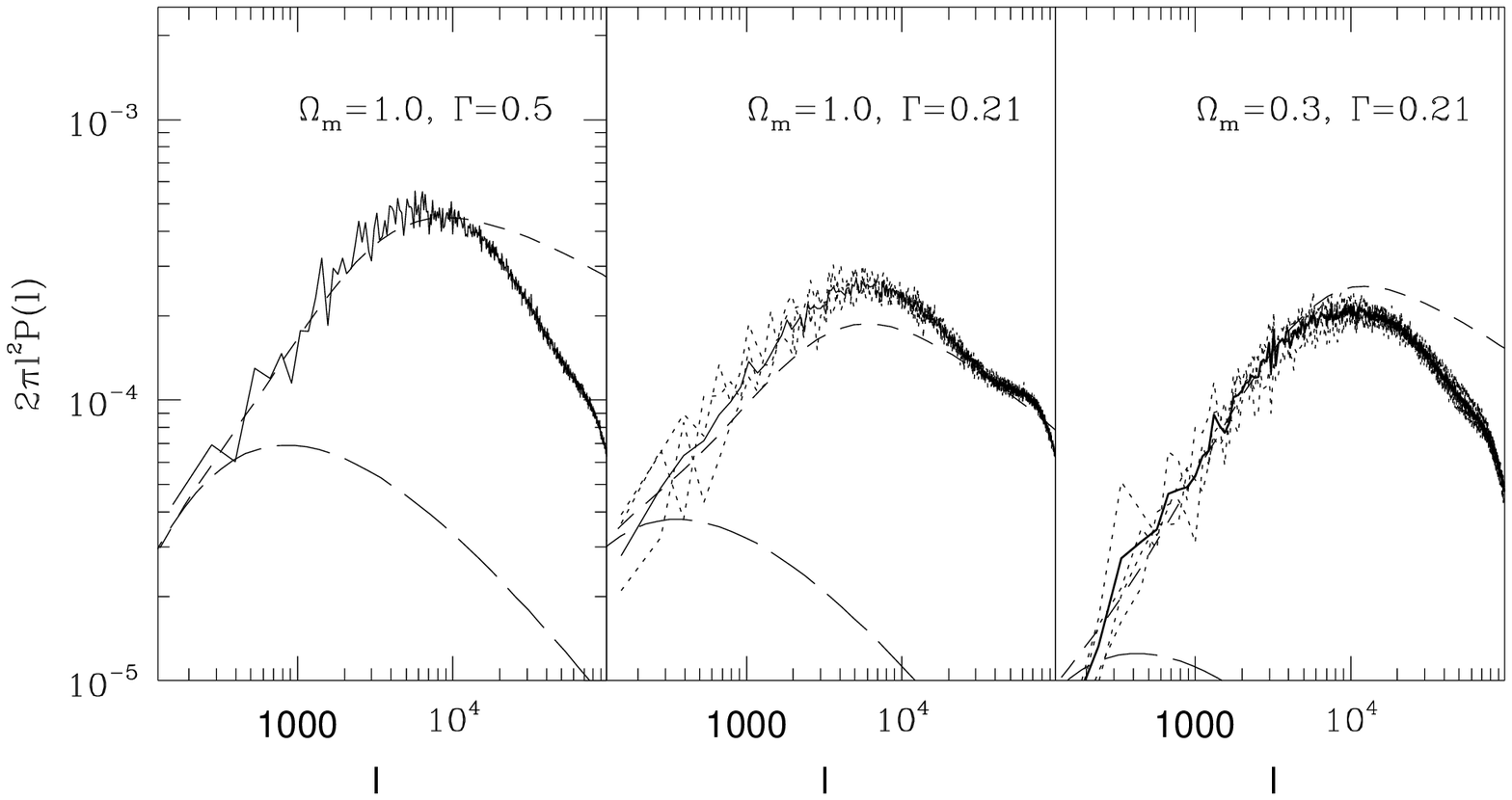}
\label{figpower}
\end{figure}

There are strong 
features due to the nonlinear clustering of the dark matter as
shown in Figure 1. Collapsed halos produce tangential patterns 
of shear around them, while quasilinear filamentary structures are 
also visible. The two cosmological models show qualitative
differences in the shear pattern, with the open model being far 
more dominated by halos and lacking irregular or filamentary structures.
Figure 2 shows the power spectrum of the convergence (essentially
identical to that of the shear) for 3 cosmological models. 
The amplitude of the convergence
is enhanced by a factor of several compared to 
linear theory on small scales, as seen from the difference between
the measured spectrum and the one predicted from linear theory. 
For a square field $2.5^{\circ}$ on a side essentially all the modes
have become  nonlinear. 
The analytical nonlinear spectrum (from Jain and Seljak 1997) 
agrees adequately (to about $20\%$)  with the measured
one over more than 2 orders of magnitude. On the smallest scales, 
the measured spectrum is suppressed due to resolution effects.
For the middle panel there are differences with the analytic 
prediction caused by too small a simulation box (85$h^{-1}$Mpc).
We ran several $PM$ simulations and found that for models with a 
lot of large scale power simulation boxes of size 200$h^{-1}$Mpc
are needed for an adequate treatment of the large scale modes.
The results from these simulations agree well with the analytic predictions.

Figure 3 shows the skewness and kurtosis of the convergence. 
The skewness on scales $\lsim 10'$ is larger than the perturbation
theory predictions shown by the dashed lines. The ratio of the
skewness for the open and $\Omega_m=1$ models shows that it is
a sensitive probe of matter density $\Omega_m$ (Bernardeau et al. 1997), 
but for a proper calibration one has to resort to N-body simulations. 
The kurtosis shows the same
trend of increasing with decreasing $\Omega_m$. However the skewness
and kurtosis are sensitive to the sampling variance
and therefore require a large survey area for a robust measurement. 

We also investigate the one-point pdf of the convergence and shear 
to assess its sensitivity to different model parameters. One example is
shown in Figure 4, where the pdf of $\kappa$ is compared with
that of the  3-dimensional overdensity $\delta=(\rho-\bar{\rho})/\bar{\rho}$. 
The comparison illustrates a key difference: whereas the pdf of
$\delta$ is fully determined by the smoothed variance (for a 
given shape of the power spectrum), that of $\kappa$
is sensitive to $\Omega_m$ as well. 
The sensitiviy of the pdf and higher moments to $\Omega_m$ arises
because the relative deflection of photons is determined by the 
projected matter density, not the relative overdensity.
Particularly robust is the difference between the mean convergence,
tracing the mean projected density, and the 
smallest negative convergence, tracing empty regions of the universe.
In this case empty voids in an open universe produce a smaller 
de-magnification relative to the mean
than in an $\Omega_m=1$ universe. This shows up as a less negative
cut-off in the pdf in Figure 4 and, unlike the moments of pdf, is
independent of the amplitude and 
shape of the power spectrum on sufficiently small scales. 
This difference at large negative
$\kappa$ as well as differences in the positive tail account for
the sensitivity of the pdf to $\Omega_m$. 
A detailed analysis of the pdf in the presence of noise
will be given in a future publication. 

\begin{figure}[p]
\vspace*{17cm}
\caption{The skewness $S_3(\theta)$ of $\kappa$ versus $\theta$ for low and
high $\Omega_m$ models. The symbols in the upper panels show the skewness 
measured from four different realizations of the ray tracing. The solid line
is the average over these realizations, while the dashed line is the
prediction from perturbation theory. The lower left panel shows the
ratio of the skewness for $\Omega_m=0.3$ and $\Omega_m=1$, again for the
different realizations and their average. The lower right panel shows
the kurtosis of $\kappa$ for the $\Omega_m=0.3$ model (triangles) and
the $\Omega_m=1$ model (circles). 
}
\includegraphics{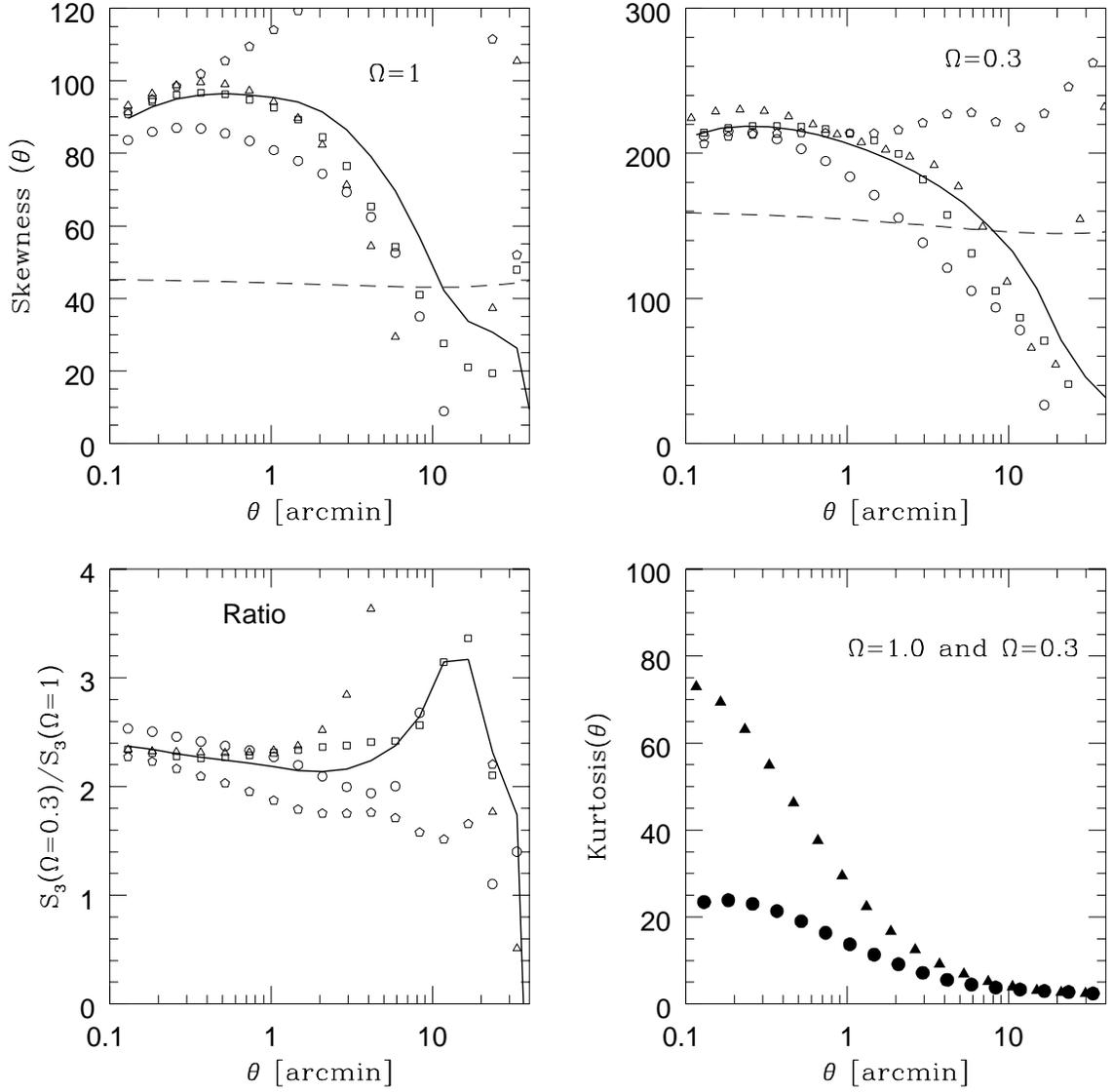}
\label{figskew}
\end{figure}

\begin{figure}[t]
\vspace*{9cm}
\caption{The probability distribution function (pdf) of $\kappa$ and of
$\delta$, the 3-dimensional over-density. The left panel show the pdf of $\kappa$
for the open (dashed) model and the Einstein-de Sitter (solid) 
model. The thick solid curve uses the same smoothing scale as the
dashed curve, $0.4'$. The thin solid curve is for a smoothing scale of
$1.6'$ which has the same rms as the dashed curve. The right panel shows
the pdf of the 3-dimensional density for the same two models, smoothed
on a scale of 2.7 $h^{-1}$ Mpc. The rms for the two models is nearly 
the same. A comparison of the two panels shows that the rms completely
dictates the pdf for $\delta$ (for the same shape of the power 
spectrum), while the pdf of $\kappa$
differs with $\Omega_m$ even if the rms is the same. 
}
\includegraphics{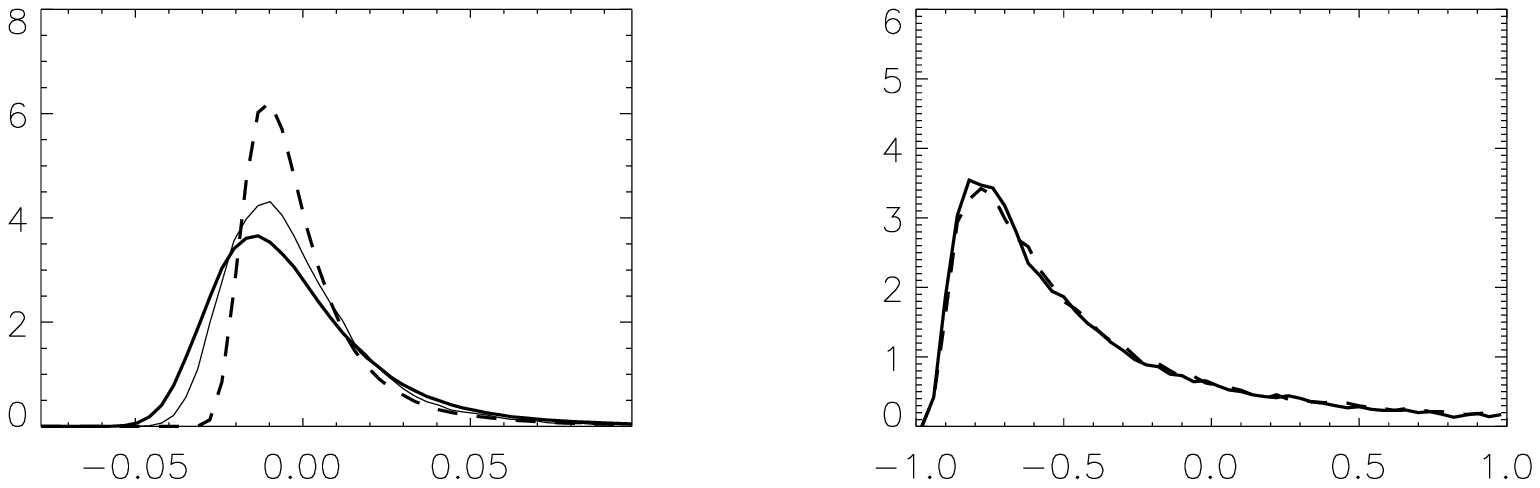}
\label{fighist}
\end{figure}

\section{Power Spectrum Estimation}
The first measure we want to extract from the observed data is the 
power spectrum in the presence of noise. On large scales (above
one degree) the data are gaussian distributed and there is a well defined
strategy to measure the power spectrum, which at the 
same time also provides the sampling and noise error estimate
(Seljak 1997). On small scales the optimal weighting of the data 
is less critical, but one needs to use the N-body simulations
to measure the sampling variance and mode-mode correlations. 

On almost all the scales from a few arcminutes to several degrees
the signal to noise in the power spectrum is much larger than unity
for a survey of degree size and flux limit of I=25-26 (Jain \& Seljak 1997).
This suggests sparse sampling (Kaiser 1997), where the survey area is
not fully covered, so that for a given amount of observing time the area
can be increased (thereby reducing the sampling variance), while
the density of galaxies is decreased (increasing the noise variance). 
If there was only noise then one could devise an
optimal sampling strategy by equating the noise and sampling variance
terms in the error matrix for the power spectrum. In practice 
however there is an additional noise term arising from the small
scale modes that are not well measured. This aliasing term (see
Seljak 1997 for its explicit form) is really a signal, but if those
modes are not sufficiently well sampled, then they cannot be 
reconstructed and therefore act as noise. This additional 
power has to be added to the usual noise. If the sampling is 
sufficiently sparse, then this term dominates over the usual noise. Since 
its amplitude depends on the amount of small scale power one has to 
measure it first before subtracting its contribution. 

This leads to the following adaptive sparse
sampling strategy: first one measures the power on small scales
using a filled survey. Once this power is measured to sufficient 
accuracy one can start to sparsely sample the sky, moving from 
smaller to larger and larger scales. At each step one minimizes
the variance, which amounts to equating the sampling variance to
the noise+aliasing variance, the later being  estimated using the 
measured power 
on smaller scales. If the amount of small scale power is small one
can afford very spase sampling, otherwise only modest sparse 
sampling can be used. In either case the formalism 
allows for an optimal sampling strategy which is adaptive in the 
sense that if the slope of the power spectrum changes as a function 
of scale so does the sampling fraction.

A $10^{\circ}\times 10^{\circ}$ field which is sparsely sampled 
gives a 2-d power spectrum shown in figure 5a. Usually one is more 
interested in a 3-d power spectrum and would therefore like to invert 
the 2-d to a 3-d spectrum. In general the 2-d power spectrum at some $l$ mode
receives contributions from a wide range of 3-d $k$ modes, with the window
peaking at $l \sim kr$, where $r$ is half the distance between 
the observer and the source galaxies. Conversely each 3-d spectral
bin estimate will receive contributions from a range of 
2-d spectral bins, 
appropriately weighted with the window so that 2-d bins close
to the peak of the window have a larger weight than those far in the tails.
In addition, one must also appropriately weight the 2-d estimators
themselves using the corresponding covariance matrix. If a given 2-d
bin has large sampling or noise variance then it should be downweighted
appropriately. The complete formalism is presented in Seljak (1997). 
Figure 5b shows for the $10^{\circ}\times 10^{\circ}$ sparsely sampled
field the 3-d reconstructed power spectrum. The reconstructed 3-d bins
are strongly correlated because of the window. This explains why the 
variance on individual estimators is smaller than in the 2-d case -- the
estimators were essentially averaged over a wider range in $k$. This 
causes no problems if the power spectrum is smooth; however, fine details
such as bumps and peaks cannot be accurately resolved with weak 
lensing data because of the smoothing. Note that we have assumed 
gaussian sampling variance, which becomes invalid on small scales 
($k>0.5hMpc^{-1}$) and will increase the errors in that regime. 

\begin{figure}[t]
\vspace*{8.3cm}
\caption{2-d power spectrum reconstruction (left) and 3-d power 
spectrum reconstruction from the 2-d power spectrum (right). Both are 
for simulated 100 square degree sparsely (1 in 4) sampled data with
a density of $2 \times 10^5$ galaxies per square degree.  The 
thick and thin solid lines on the left are for the nonlinear and 
linear power spectrum, respectively, while the dashed line is the 
power spectrum with $45^{\circ}$ rotation. The latter should vanish
in a perfect survey
if weak lensing signal is generated by gravity. If the sampling is 
sparse aliasing mixes power into this combination. The amount of 
aliasing can be estimated once the true power is measured, in 
which case on can verify the assumption of weak lensing signal 
generated by gravity.
On the right the solid line 
is the nonlinear 3-d power spectrum. A survey of this size can 
reconstruct the power spectrum up to the turnover scale.}
\includegraphics{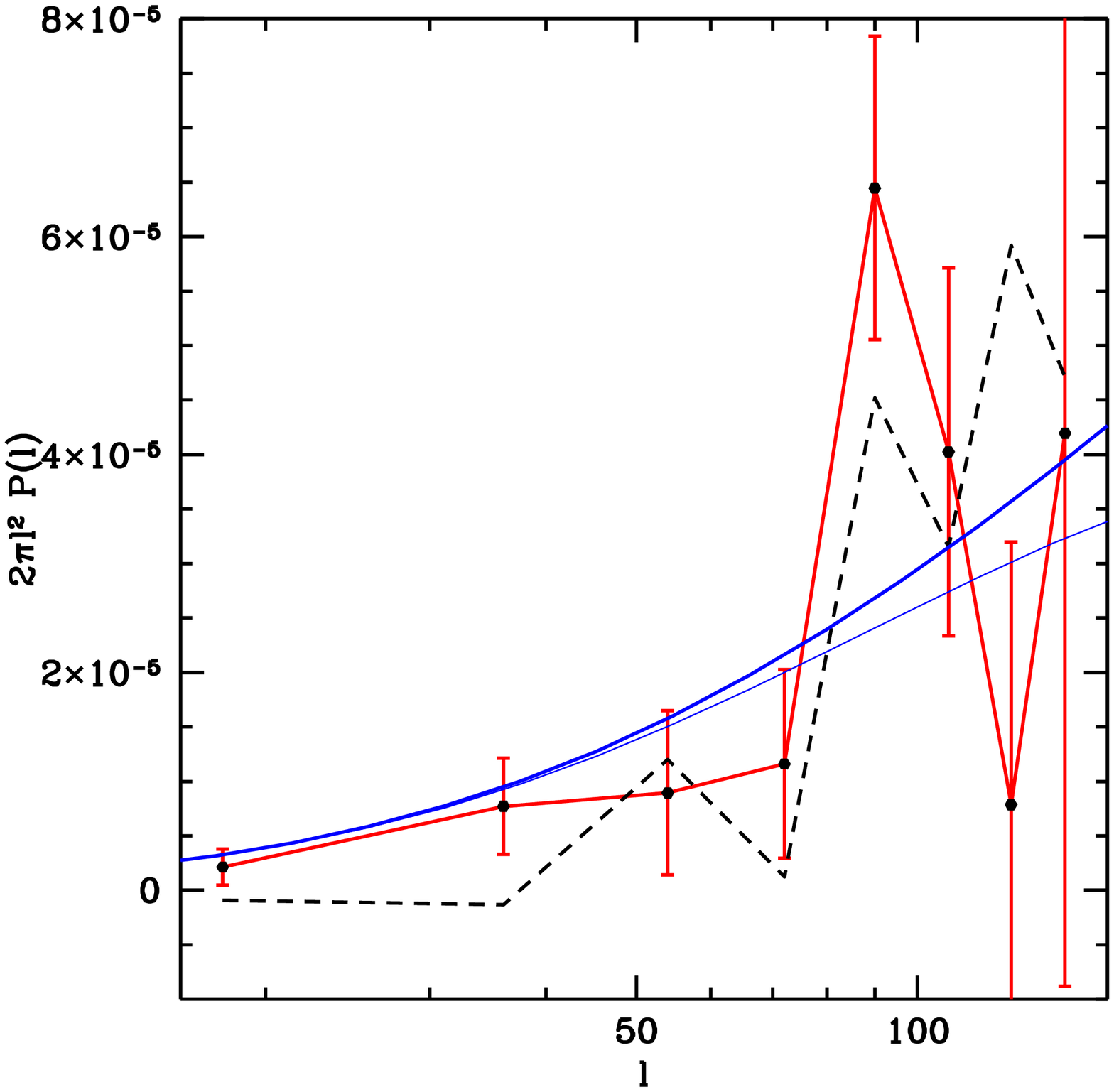}
\includegraphics{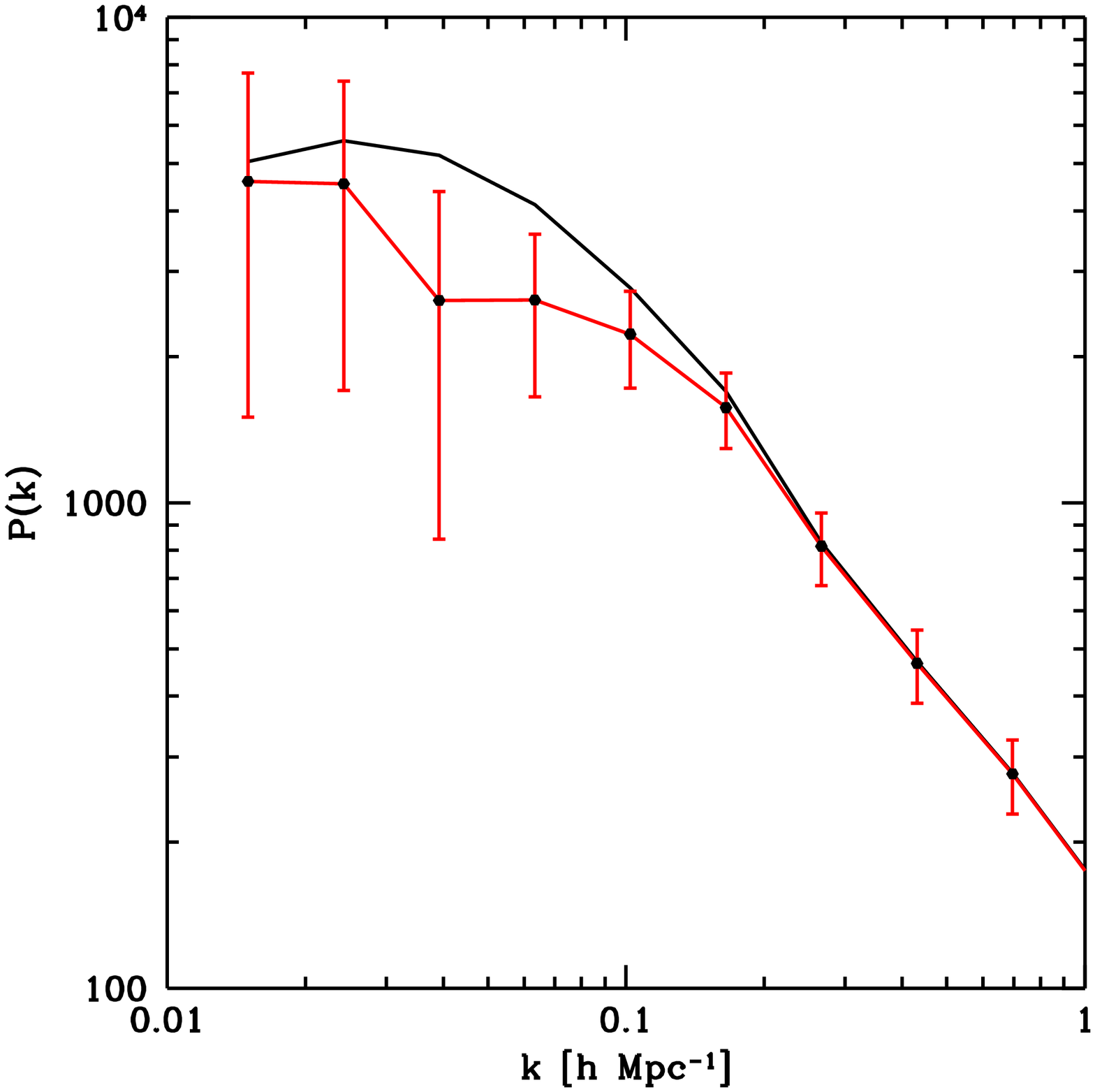}
\end{figure}

\section*{Acknowledgments}

The high resolution simulations in this paper were carried out
using codes made available by the Virgo consortium. We thank 
Joerg Colberg for help in accessing this data
and Ed Bertschinger for making available his $PM$ N-body code.
We are grateful to Matthias Bartelmann, Ue-Li Pen, Peter Schneider and
Alex Szalay for useful discussions.

\end{document}